\documentclass[aps,twocolumn]{revtex4}
\usepackage{graphicx}
\input{psfig}

\begin{document}

\title{Experimental demonstration of an analytic method for image \\ 
  reconstruction in optical tomography with large data sets}

\author{Zheng-Min Wang, George Y. Panasyuk, Vadim A. Markel and John C. Schotland}

\affiliation{Departments of Bioengineering and Radiology \\ University of Pennsylvania, Philadelphia, PA 19104}

\date{\today}

\begin{abstract}
  We report the first experimental test of an analytic image
  reconstruction algorithm for optical tomography with large data
  sets. Using a continuous-wave optical tomography system with $10^8$
  source-detector pairs, we demonstrate the reconstruction of an
  absorption image of a phantom consisting of a highly-scattering
  medium with absorbing inhomogeneities.
\end{abstract}
 
\maketitle

Optical tomography (OT) is a biomedical imaging modality that utilizes
diffuse light as a probe of tissue structure and
function\cite{Gibson_2005}. Clinical applications include imaging of
breast disease and functional neuroimaging. The physical problem that
is considered is to reconstruct the optical properties of an
inhomogenous medium from measurements taken on its surface. In a
typical experiment, optical fibers are used for illumination and
detection of the transmitted
light~\cite{Schmidt_2000,Mcbride_2001,Colak_1999}. The number of
measurements (source-detector pairs) which can be obtained, in
practice, varies between $10^2-10^4$. A recently proposed alternative
to fiber-based experiments is to employ a narrow incident beam for
illumination. The beam can be scanned over the surface of the medium
while a lens-coupled CCD detects the transmitted light. Using such a
``noncontact'' method, it is possible to avoid many of the technical
difficulties which arise due to fiber-sample
interactions~\cite{Schulz_2003,Ripoll_2004,Turner_2005,Cuccia_2005}.
In addition, extremely large data sets of approximately $10^8-10^{10}$
measurements can readily be obtained. Data sets of this size have the
potential to vastly improve the quality of reconstructed images in OT.

The reconstruction of images from large data sets is an extremely challenging problem due to the high computational complexity of numerical approaches to the inverse problem in OT. To address this challenge, we have developed analytic methods to solve the inverse problem~\cite{Schotland_1997,Markel_2002-2,Markel_2004}. These methods lead to a dramatic reduction in computational complexity and have been applied in numerical simulations to data sets as large as $10^{10}$ measurements~\cite{Markel_2002-2}. In this Letter, we report the first experimental test of an analytic image reconstruction method. By employing a noncontact OT system with $10^8$ source-detector pairs, we reconstruct the optical absorption of a highly-scattering medium. The results demonstrate the feasibility of image reconstruction for OT with large data sets.

We begin by considering the propagation of diffuse light. The density
of electromagnetic energy $u({\bf r})$ in an absorbing medium obeys the
diffusion equation

\begin{equation}
\label{diff_eqn}
-D\nabla^2 u({\bf r}) + \alpha({\bf r})u({\bf r}) = S({\bf r}) \ ,
\end{equation}

\noindent
where $\alpha({\bf r})$ is the absorption coefficient, $S({\bf r})$ is
the power density of a continuous wave source, and $D$ is the
diffusion constant. The energy density also obeys the boundary
condition $u+\ell \hat{\bf n}\cdot\nabla u=0$ on the surface bounding
the medium, where $\hat{\bf n}$ is the unit outward normal and $\ell$
is the extrapolation length~\cite{Markel_2002-1}. The relative
intensity measured by a point detector at ${\bf r}_2$ due to a point
source at ${\bf r}_1$ is given, within the accuracy of the first Rytov
approximation, by the integral equation

\begin{equation}
\label{rytov}
\phi({\bf r}_1,{\bf r}_2) = \int d^3r G({\bf r}_1,{\bf r})G({\bf r},{\bf r}_2)\delta\alpha({\bf r}) \ ,
\end{equation}

\noindent
where the source and detector are oriented in the inward and outward
normal directions, respectively~\cite{Markel_2004}.  Here
$\delta\alpha({\bf r})=\alpha({\bf r})-\alpha_0$ denotes the spatial
fluctuations in $\alpha({\bf r})$ relative to a reference medium with
absorption $\alpha_0$, $G$ is the Green's function for
Eq.~(\ref{diff_eqn}) with $\alpha=\alpha_0$, and the data function
$\phi$ is defined by
$\phi({\bf r}_1,{\bf r}_2)=-G({\bf r}_1,{\bf r}_2)\ln(I({\bf r}_1,{\bf r}_2)/I_0({\bf r}_1,{\bf r}_2))$, where
$I({\bf r}_1,{\bf r}_2)$ denotes the intensity in the medium and
$I_0({\bf r}_1,{\bf r}_2)$ is the intensity in the reference medium. Note that
the intensity is related to the Green's function by the expression

\begin{equation}
\label{int}
I({\bf r}_1,{\bf r}_2) = {cS_0\over 4\pi}\left(1+{\ell^*\over\ell}\right)^2G({\bf r}_1,{\bf r}_2) \ ,
\end{equation}

\noindent
where $S_0$ is the source power and the transport mean free path
$\ell^*$ is related to the diffusion coefficient by $D=1/3 c\ell^*$.

We have constructed a noncontact OT system to test the analytic method
of image reconstruction. A schematic of the instrument is shown in
Fig.~1. The source is a continuous-wave stabilized diode laser
(DL7140-201, Thorlabs) operating at a wavelength of 785 nm with an
output power of 70 mW. The laser output is divided into two beams by a
beam splitter. The reflected beam is incident on a power meter which
monitors the stability of the laser intensity. The transmitted beam
passes through a lens onto a pair of galvanometer-controlled mirrors
(SCA 750, Lasesys). The mirrors are used to scan the beam, which has a
focal spot size of 200 $\mu$m, in a raster fashion over the surface of
the sample. After propagating through the sample, the transmitted
light passes through a band-pass interference filter (10LF20-780,
Newport) and is imaged onto a front illuminated thermoelectric-cooled
16-bit CCD array (DV435, Andor Technology) using a 23 mm/$f$1.4 lens.
A mechanical shutter is placed in front of the CCD to reduce artifacts
associated with frame transfer within the CCD chip. A pulse generator
with digital delay is used to trigger and synchronize the CCD, the
shutter and the position of the beam.

\centerline{\centerline{\psfig{file=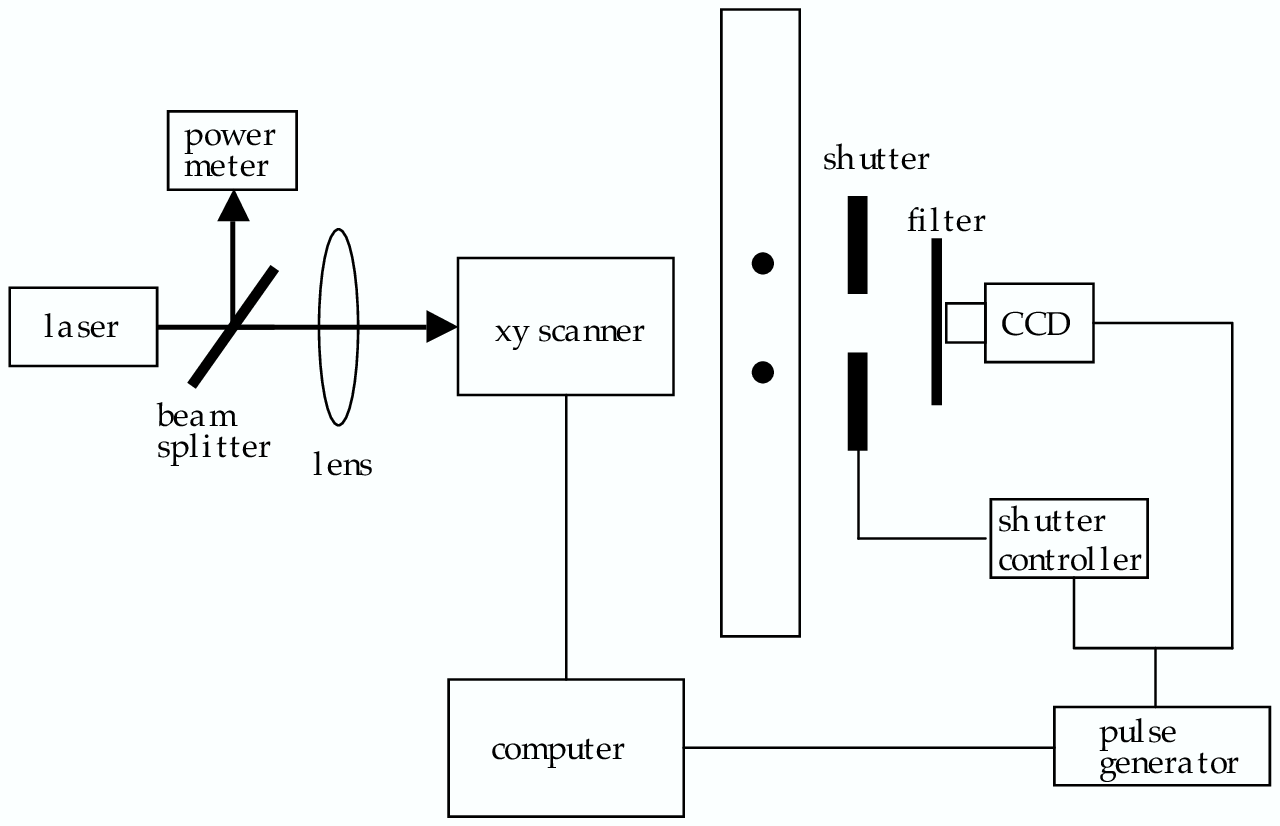,width=8.5cm,bbllx=60bp,bblly=280bp,bburx=500bp,bbury=575bp,clip=t}}}
{\small Fig.~1. Schematic of the noncontact optical tomography system.}\\

The sample chamber is a rectangular box of depth 5 cm with square
faces of area $50\times50$ cm$^2$ constructed of clear acrylic sheets.
The beam is scanned on one face of the sample and the opposite face is
imaged by the CCD. The chamber is placed equidistantly from the CCD
and the laser source along the optical axis at a distance of 110 cm.
The chamber is filled with a scattering medium which consists of a
suspension of 1\% Intralipid in water in which absorbing objects may
be suspended.

A tomographic data set is acquired by raster scanning the beam over a
$29\times29$ square lattice with a lattice spacing of 0.5 cm. This
yields 841 source positions within a $14\times14$ cm$^2$ area centered
on the optical axis. For each source, a $429\times429$ pixel region of
interest is read out from the CCD. This results in 184,041 detectors
arranged in a square lattice with an effective lattice spacing
equivalent to $0.065$ cm and all detectors located within a $28 \times
28$ cm$^2$ area centered on the optical axis. Thus a data set of $1.5
\times 10^8$ source-detector pairs is acquired.

The inverse problem in OT consists of reconstructing $\delta\alpha$
from measurements of $\phi$. In this Letter, we consider the inversion
of the integral equation (\ref{rytov}) in the slab measurement
geometry. The approach taken is to construct the singular value
decomposition of the integral operator whose kernel is defined by
(\ref{rytov}) and to use this result to obtain the pseudoinverse
solution to (\ref{rytov}).

The starting point for this development is to consider the lattice Fourier transform of the sampled data function which is defined by
\begin{equation}
\label{fourier}
\widetilde\phi({\bf q}_1,{\bf q}_2) = \sum_{{\bf r}_1,{\bf
    r}_2}\exp\left[i({\bf q}_1\cdot{\bf r}_1 + {\bf q}_2\cdot{\bf
    r}_2)\right] \phi({\bf r}_1,{\bf r}_2) \ ,
\end{equation}
where the sum is carried out over the square lattices of sources and
detectors with lattice spacings $h_1$ and $h_2$, respectively. The
wave vectors ${\bf q}_1$ and ${\bf q}_2$ belong to the first Brillouin zones of
the corresponding lattices, denoted FBZ($h_1$) and FBZ($h_2$). It can
then be shown that the pseudoinverse solution to the integral equation
(\ref{rytov}) is given by the inversion formula
\begin{equation}
\label{inv_formula}
\delta\alpha({\bf r}) = \int_{{\rm FBZ}(h_1)} d^2q \int_{{\rm FBZ}(h_2)}d^2p K({\bf r};{\bf q},{\bf p})\widetilde\phi({\bf q}-{\bf p},{\bf p}) \ ,
\end{equation}
where the kernel $K$ is defined in Ref.~\cite{Markel_2004}. Several
aspects of Eq.~(\ref{inv_formula}) are important to note. First, the
transverse spatial resolution of reconstructed images is determined by
the spatial frequency of sampling of the data function with respect to
both source and detector coordinates. As a consequence, a large number
of source-detector pairs is required to achieve the highest possible
spatial resolution. It can be seen that when the source and detector
lattices have equal spacing, the theoretical limit of transverse
resolution is given by the lattice spacing. When the source and
detector lattice spacings are different, as is the case in the
experiment reported here (where $h_1=0.5$ cm and $h_2=0.065$ cm), the
resolution of reconstructed images is controlled by the larger lattice
spacing (lower spatial frequency). Second, the inverse problem in OT
is evidently overdetermined. In addition, it is highly ill-posed. As a
result, it can be said that large data sets allow for averaging the
data function in such a way that the sensitivity to noise in the
inverse problem is partially ameliorated. Finally, numerical
implementation of (\ref{inv_formula}) requires replacing the integrals
over $d^2q$ and $d^2p$ by sums over a finite set of wavevectors. In
practice, we find that integration over $d^2q$ can be carried out with
a step size $\Delta q =0.07$ cm$^{-1}$ and 14,641 integration points
while the integration over $d^2p$ requires a step size $\Delta p=1/2
\Delta q $ and 1,296 integration points. Thus a total of
$1.9\times10^7$ Fourier components of the data are used in the
reconstruction.

The first step in the reconstruction of tomographic images is to
measure the reference intensity $I_0$ for each source-detector pair.
By fitting this data in the spatial frequency domain to (\ref{int})
with $\alpha=\alpha_0$ we obtain the diffuse wavenumber
$k_0=\sqrt{\alpha_0/D}=0.58$ cm$^{-1}$ and the extrapolation length
$\ell= 0.7$ cm. Note that these parameters define the diffusion
Green's function $G$ in the slab geometry~\cite{Markel_2002-1} and
that $\alpha_0$ and $D$ cannot be separately determined from a
continuous-wave measurement at a single wavelength.  Next, the object
to be imaged is placed in the sample chamber and the intensity $I$ for
each source-detector pair is measured. In Fig.~2 we show the
reconstruction of a pair of black metal balls. The balls have a
diameter of 8 mm and were suspended in the midplane of the sample
chamber at a constant height with a separation of 3.2 cm. Tomographic
images were reconstructed with a $15\times15$ cm$^2$ field of view
using $230\times230$ pixels per image with a separation between the
slices of 0.26 cm. It can be seen in the central slice, which is
equidistant from the source and detector planes, that the balls are
well resolved.  The shallower and deeper slices show that the balls
remain well resolved but with a smaller diameter, as expected. Fig.~3
is a plot of $\delta\alpha/\alpha_0$ along the line passing through
the centers of both balls in the central slice. The distance between
the peaks is 3.3 cm in close agreement with the measured separation of
the balls. The FWHM of the peaks is 1.1 cm which slightly
overestimates the diameter of the balls. The FWHM of the peaks in the
depth direction is 1.5 cm (graph not shown).  It is important to note
that the reconstructed contrast in $\delta\alpha$ is not expected to
be quantitative due to the possible breakdown of (\ref{rytov}) in the
interior of the strongly absorbing balls. Interestingly, however, the
shape and volume of the spherical absorbers is recovered well.

\centerline{\psfig{file=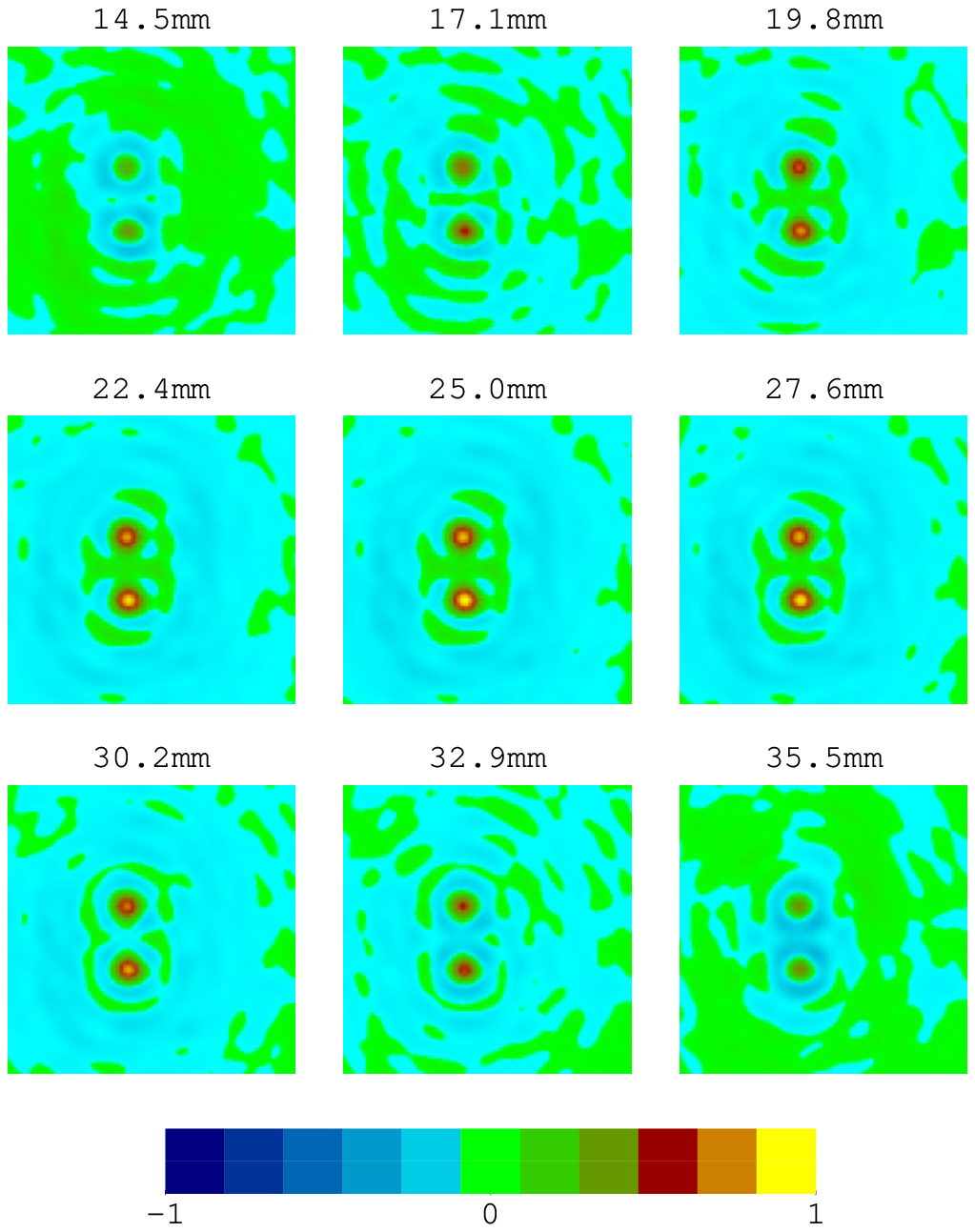,width=9.0cm,bbllx=90bp,bblly=320bp,bburx=450bp,bbury=720bp,clip=t}}
{\small Fig.~2. Reconstructions of $\delta\alpha$ for the two-ball phantom plotted on a linear color scale. The distance of each slice from the plane of sources is indicated. All images are normalized to the maximum of the central slice.}\\

In conclusion, we have demonstrated the feasibility of analytic
methods for image reconstruction in OT with large data sets. We are
currently conducting further studies to assess the effects of
absorption contrast on image resolution. In addition, the recent
availability of CCDs with faster data acquisition will allow the
collection of data sets with greater numbers of sources, leading to
improvements in spatial resolution. We expect that with further
technological advances, resolution consistent with the results of
numerical simulations~\cite{Markel_2002-2} will be achieved.

This research was funded by the NIH under the grants P41RR02305 and
R21EB004524. Support from the Whitaker Foundation is also gratefully
acknowledged.

\centerline{\psfig{file=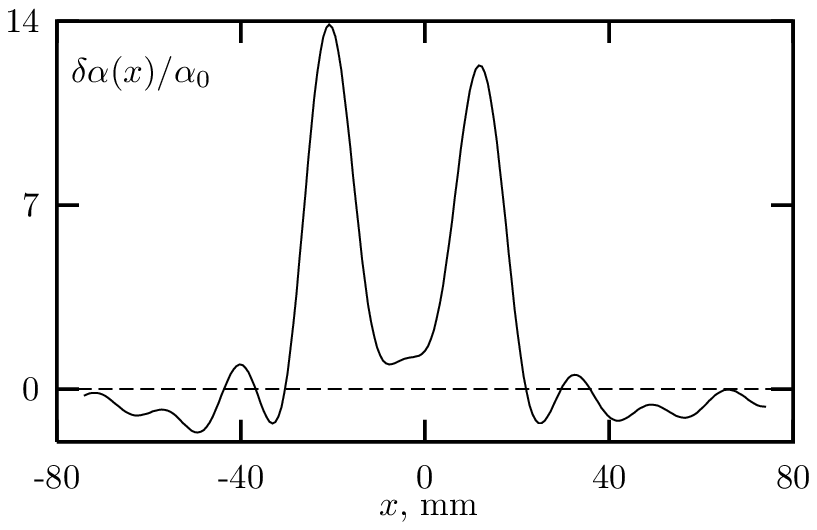,width=8.5cm,bbllx=175bp,bblly=570bp,bburx=421bp,bbury=750bp,clip=t}} 
{\small Fig.~3. A one-dimensional profile of the reconstructed
  absorption along the line passing through the centers of the balls
  in the central slice.}\\

\end{document}